\documentclass[11pt,twoside]{article}

\usepackage{asp2006}
\usepackage{graphicx}
\usepackage{lscape}

\begin{document}
\title{Observational links between AGN evolution and galaxy growth}   %%% Fill in title
\author{L. Miller}   %%% Fill in author names
\affil{Dept. of Physics, Oxford University, Oxford OX1 3RH, U.K.}    %%% Fill in author affiliations

\begin{abstract} %%% Abstract to run on from here.
There is growing interest in the possible link between
the growth of supermassive black holes and the effect of
feedback from them on galaxy growth.  There are three 
areas of significant uncertainty: (i) the physics of the feedback; 
(ii) the prevalence and effectiveness of feedback; 
(iii) the link between the growth of black holes and their hosts.  
The 2QZ optical QSO survey indicates that luminous QSOs
are relatively short-lived, and it has recently been shown that 
the observed bolometric luminosity density from all AGN 
and its evolution can be reproduced if black holes grew coevally with their
galaxies, implying but not requiring a causal link between galaxy growth
and black hole growth.  At low redshifts there is some evidence that
black hole and galaxy growth are starting to decouple.
\end{abstract}

%%% MAIN BODY OF TEXT GOES HERE. CONSULT "INSTRUCTIONS FOR AUTHORS USING
%%% LATEX2E MARKUP", SECTIONS 2.3-2.6 FOR HELP WITH EQUATIONS, FIGURES,
%%% AND TABLES.

%\section{}   %%% Top level section head (remove "%" symbol)
%\subsection{}   %%% Second level section head (remove "%" symbol)
%\subsubsection{}   %%% Lowest level section head (remove "%" symbol)
%\section*{}    %%% Unnumbered top level section head (remove "%" symbol)
%\subsection*{}   %%% Unnumbered second level section head (remove "%" symbol)

\section{Introduction}
For many years the active galaxy phenomenon was regarded as an interesting sideline
in astrophysics that related solely to our understanding of black holes.  However,
with the recognition that every massive galaxy in the local universe has a 
supermassive black hole at its heart 
\citep{magorrian98, richstone98, ferrarese00, gebhardt00}
we now recognise that black hole formation is an integral part of galaxy formation,
with relationships between black hole and host galaxy that are consistent for
galaxies with or without AGN \citep{onken04, barth05}.
Moreover, it now seems possible that feedback from black hole growth may have a
significant effect in shutting off star formation and defining the colour-luminosity 
distribution of massive galaxies \citep{benson03, croton06, bower06, delucia06}.
However, our understanding of both the physics and the prevalence of AGN feedback is
extremely limited, as reviewed in the next section.  We also have only a partial
understanding of the cosmological dependence of black hole growth within galaxies,
yet if we wish to understand AGN feedback in the cosmological context this is
surely a crucial aspect of the problem.
In later sections of this article I discuss one particular scenario for black hole
growth, namely that black holes and galaxies and their dark halos grow coevally,
and compare with observation.

\section{AGN feedback - fact or fiction?}
The physics of supermassive black hole growth in galaxies and its cosmological
evolution has recently been receiving wide attention because of the possible
effects of AGN feedback on the formation of the host galaxy itself \citep[e.g.][]{dimatteo05}.
Since we now recognise that every massive galaxy (at least) has a black hole, such
feedback may be an integral part of the general galaxy-formation process.
\citet{benson03} have argued that something like AGN feedback is required to shut
down star formation in the most massive galaxies in order to recreate the observed
galaxy mass function, and it has been argued that such feedback
is required to reproduce the observed colour-luminosity distribution of galaxies
\citep{bower06, croton06, delucia06}.
However, the feedback process has to operate in every galaxy and in
the relatively recent universe,
so rather than feedback from the luminous QSO or AGN phase of black hole growth,
these authors envisage a ``radio mode'' in which a radiatively-quiet, 
but kinetically-powerful, jet or outflow provides the feedback in the recent
universe.  

It has long been suggested that radio sources may provide significant heat input to
cluster gas \citep[e.g.][]{miller88,pedlar88,pedlar90} but this idea wasn't really taken
seriously until the discovery both of a lack of cooling gas in ``cooling flow'' clusters
and of a possible link with radio structures in massive clusters such as Perseus
(\citealt{peterson_fabian06} and references therein).  But most galaxies at the present
day don't have powerful radio sources associated with them, so we need to consider
the evidence for ``quiet'' outflows in nearby galaxies.

Nearby Seyfert active galaxies do have low density outflowing gas in their nuclear regions,
but so far these do not appear energetically significant: in NGC\,5548 for example the
mass outflow rate is $\dot{M} \sim 0.3$\,M$_{\odot}$\,year$^{-1}$ 
with a maximum outflow velocity $\sim 1000$\,km\,s$^{-1}$ \citep{steenbrugge05}, 
so it seems unlikely that this
amount of feedback would significantly affect the host galaxy.  Evidence for 
higher velocities and therefore
significantly higher mass and momentum outflow rates has been seen in a few QSOs
\citep[e.g.][]{pounds03}.  High velocity QSO outflows are also seen in UV absorption lines,
occurring in $\sim 26$\,percent of QSOs \citep{trump06}, and it has been supposed that
these might be associated with an energetic disc wind \citep[e.g.][]{proga_kallman04},
but if such outflows are restricted to the rare (at $z=0$) luminous QSO phase of
black hole activity they won't do the job.  But kinetic jet output may be more significant
than outflowing winds \citep[e.g.][]{omma04}, and radiatively-inefficient flows
may produce such outflows \citep{narayan94, ho02}, or indeed rotating black holes may 
produce output electrodynamically \citep{blandford_znajek77,reynolds06} 
perhaps independently of
visible radiation.  But a requirement of all these pictures is that, regardless of
whether or not we can see it, the source of the feedback energy is gravitational - accretion
of matter onto the black hole.  Despite many attempts over the past four decades,
we still have limited understanding
of how black holes form in galaxies and why their luminous accretion phase shows such
strong cosmological evolution.
In the remainder of this article we discuss what we know about
black hole accretion and its cosmological history, and in particular discuss the
hypothesis that black holes and galaxies grew coevally.

\section{QSO and AGN evolution}
Twenty years ago a striking picture emerged of the cosmological evolution of luminous QSOs:
they had an optical luminosity function that had a broken power-law form and that
evolved steadily to lower luminosity at lower redshifts without changing in normalisation
\citep{marshall83, boyle88}. A natural interpretation of this was that supermassive black
holes formed relatively early in the universe, and that accretion onto them steadily
declined with time to produce the apparent ``pure luminosity evolution''.  More recently
however this picture has fallen into disfavour, for two reasons.  

First, it is widely
believed that dark matter haloes and their galaxies have grown hierarchically, with massive
structures continuing to build up in the relatively recent universe.  It is often hypothesised
that QSOs are triggered by mergers between galaxies in such a hierarchical universe, and
models seeking a cosmological explanation for QSO evolution have been based on a merger-driven
build-up of black holes \citep[e.g.][]{kauffmann_haehnelt00}.  This link to galaxy build-up
would explain the qualitative similarity in the cosmological evolution of star formation 
and nuclear black hole activity \citep[e.g.][]{dunlop97, boyle98, percival_miller99}.

Second, a more detailed look at the luminosity function has revealed departures from 
pure luminosity evolution, with evolution in the slope of the luminosity function at both
bright and faint magnitudes \citep{hewett93, goldschmidt98, hopkins06a, fan06}.
More significantly, it seems that the peak in the comoving space density of AGN/QSOs of
a given luminosity shifts systematically to lower redshifts with decreasing luminosity
(\citealt{steffen03}, 
\citealt{cowie03}, 
\citealt{ueda03}, 
\citealt{zheng04}, 
\citealt{barger05}, 
\citealt{lafranca05}, 
\citealt{hasinger05}, and 
\citealt{hopkins06b}),
a phenomenon that has become known as cosmic downsizing.  The downsizing is often
interpreted as reflecting an increasing prevalence of lower-mass black hole growth
with decreasing redshift (but note that, unlike the case of galaxy downsizing, 
the black hole masses are not well determined at the redshifts covered by the above
surveys, so this interpretation can only be regarded as preliminary).

Despite these concerns, the basic picture of twenty years ago must nonetheless be correct:
the comoving irreducible mass in black holes cannot decrease with cosmic time, so the
decrease observed in the integrated AGN luminosity density (and in the differential luminosity
function) must indeed arise from a mean accretion rate that decreases with cosmic time.
The departures from pure luminosity evolution are then most likely indicating that we are
not observing a single fixed population of long-lived black holes but rather the 
statistical changes in a relatively (compared to the Hubble time) short-lived
population.

Confirmation that luminous QSO lifetimes are shorter than the Hubble time comes from
measurement of their clustering properties: the lack of clustering growth to
lower redshift shows that QSOs cannot be a long-lived population of objects
(\citealt{croom05}, these proceedings).  The upper limit on their mean lifetime
is redshift dependent and does depend on the bias model assumed for QSOs, but
a reasonable model yields $2\sigma$ limits on their existence
within the 2QZ survey of $< 2$\,Gyr and $< 1$\,Gyr at $z=1$ and $z=2$ respectively.

Hence the picture that now emerges is that AGN evolution must be caused by a decline
in overall accretion rate onto black holes, but with a characteristic timescale that
indicates a cosmological influence on a population of objects whose active lives are
short. We can measure the characteristic timescale for the QSO population change 
from the optical luminosity function.  If the characteristic break luminosity,
$L^{\star}$ varies as $(1+z)^{\gamma}$, then the characteristic timescale 
may be expressed as 
$$
\tau \equiv \frac{L^{\star}}{|dL^{\star}/dt|} = \frac{1+z}{\gamma dz/dt} = \frac{1}{\gamma H(z)}
$$
where for the optical LF, $\gamma \simeq 3$.  In the next section we see whether
a cosmological origin for this timescale can be identified.

\section{The dark halo accretion rate}
The growth of dark halos can be calculated within the framework of hierarchical
growth using the extended Press-Schechter \citep{lc93,lc94}
approach \citep{miller06}. The timescale for growth of halos of mass $M_{\rm H}$ is 
$$
\tau_{\rm H} \equiv \frac{M_{\rm H}}{\left\langle dM_{\rm H}/dt \right\rangle} = \frac{1}{f(M_{\rm H}) \left | d\delta_c/dt \right |},
$$
where $f(M_{\rm H})$ is a slowly-varying function of mass, of order unity, 
and where $\delta_c$ is the usual
Press-Schechter redshift-dependent critical overdensity for collapse (see \citealt{miller06}
for full details).  For an Einstein-de-Sitter universe this has a simple form,
$$
\tau_{\rm H, EdS} = \frac{1}{1.68 f(M_{\rm H}) H(z) (1+z)},
$$
which has a similar value at $z \sim 1$ to that observed in the QSO luminosity
function.  The timescale is insensitive to the choice of cosmology.  It is tempting 
then to suppose that the timescales for black hole growth and dark halo (and hence galaxy)
growth are comparable and perhaps related.

\section{Coeval evolution of black holes and their hosts}
One of the striking features of the black hole/bulge $M-\sigma$ relation in
the local universe is its remarkably small scatter, with an intrinsic dispersion
no larger than $\sim 0.3$\,dex \citep{tremaine02}.  It seems likely that feedback
between black hole and galaxy growth is required to produce such a tight relation
\citep[e.g.][]{king05}, but whatever the mechanism it seems most natural to suppose
that the black hole has acquired its mass at the same time as the galaxy has acquired
its mass: i.e., that black holes and galaxies have grown coevally.  \citet{miller06}
have investigated the hypothesis that the timescales for black hole and galaxy
growth are the same and show the same cosmological dependence.  The hypothesis
is that, {\em averaged across all galaxies at any given cosmological epoch}, 
\begin{equation}
\tau_{\rm BH}(z) \equiv \frac{M_{\rm BH}}{\left\langle dM_{\rm BH}/dt \right\rangle}
= \frac{M_{\rm H}}{\left\langle dM_{\rm H}/dt \right\rangle}
\equiv \tau_{\rm H}(z),
\label{eqn:pce}
\end{equation}
where $M_{\rm BH}$ and $M_{\rm H}$ are the mass of a black hole and its dark
halo respectively. We call this
``Pure Coeval Evolution'' (PCE).  The hypothesis does not require
there to be a direct causal link between the two, but it may be that feedback processes
drive the system towards this behaviour.

There are two key predictions of this hypothesis. First, the mean Eddington ratio 
of black hole accretion, averaged over all galaxies, should
rise dramatically to high redshifts, as shown in Babi\'{c} et al. (these proceedings).
Note that at any given epoch there is expected to be a wide range of individual
Eddington ratios:  galaxies with the highest values, close to unity, would be those
recognised as AGN or QSOs.  If Eddington ratios do have an upper bound around unity,
we do not expect the Eddington ratio of the most luminous
AGN to show much cosmological evolution \citep[see][]{kollmeier06}.  Less luminous
AGN and normal galaxies may show evidence for such evolution, however
\citep{netzer06}. Averaged over all galaxies, not just AGN, 
the mean Eddington ratio should increase at higher redshift, implying a greater
prevalence of visibly accreting black holes at higher $z$.

\begin{figure}
 \begin{center}
  \resizebox{10cm}{!}{ 
  \rotatebox{270}{
  \includegraphics{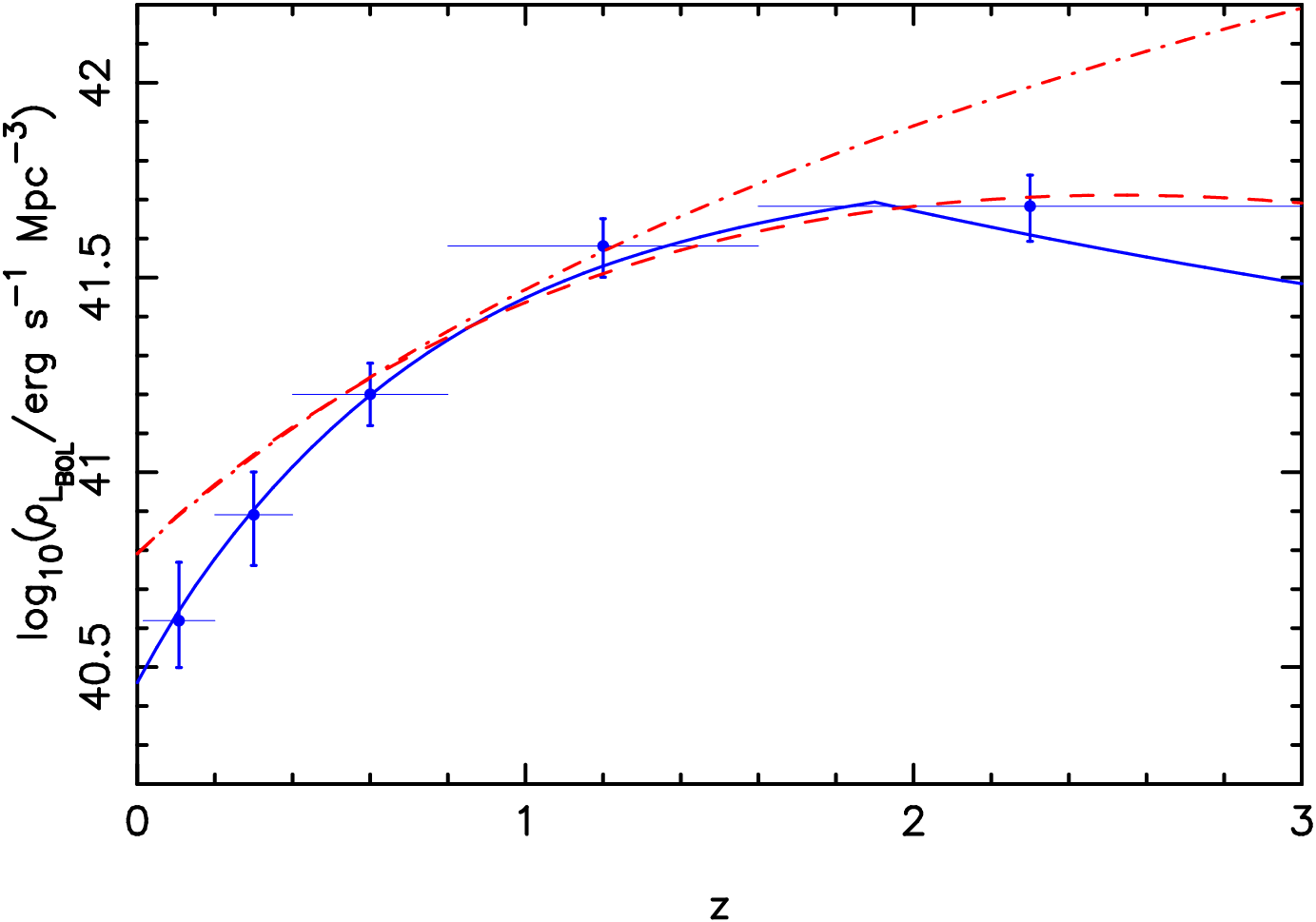}
  }}
 \end{center}
  \caption{
The AGN bolometric luminosity density deduced from the best-fit model of
\citet{ueda03}, integrating over the range 
$10^{40}<L_X<10^{48}$\,erg\,s$^{-1}$ and applying the bolometric
correction of \citet{marconi04} and correction for Compton-thick
AGN of \citeauthor{ueda03} 
(solid curve).  Also shown are uncertainties estimated
from refitting to the binned data of \citeauthor{ueda03} (points with error
bars: horizontal bars indicate the range of redshifts included
in each point).
The luminosity density expected in PCE is shown for two cases:
(i) no evolution in the comoving black hole mass density (dot-dashed upper
curve); (ii) evolution in the comoving black hole mass density that tracks
the evolution of massive dark halos with $M_{\rm H}>10^{11.5}$\,M$_{\odot}$
(dashed curve).  
Both curves assume average radiative efficiency $\langle\epsilon\rangle=0.04$.  
}
\label{fig1}
\end{figure}

Second, we can predict the expected bolometric output from accreting black holes,
as follows.  The integrated bolometric luminosity density $\rho_{\rm L}$
produced by AGN depends on the
number density of black holes, on the mean accretion rate onto those, and on the
mean radiative efficiency 
$\left\langle\epsilon\right\rangle$.  If we adopt the simplest assumption, that
equation\,\ref{eqn:pce} applies on average to black holes of all masses,
and if we approximate the function $f(M_{\rm H})$ as being independent of mass 
\citep[see][]{miller06} then we can write
\begin{eqnarray}
\nonumber
\left\langle\rho_{\rm L}\right\rangle 
&\simeq &
c^2 \rho_{\rm BH} 
\left\langle 
\frac{\epsilon}{\left(1-\epsilon\right)M_{\rm BH}}
\frac{dM_{\rm BH}}{dt} 
\right\rangle\\
&\simeq &
c^2 \rho_{\rm BH} 
\left\langle 
\frac{\epsilon f\left(M_{\rm H}\right)}{\left(1-\epsilon\right)} 
\right\rangle
\left |
\frac{d\delta_{c}}{dt} 
\right |,
\end{eqnarray}
where $\rho_{\rm BH}$ is the cosmic black hole mass density.  Here we adopt
the value for  $\rho_{\rm BH}$ at $z=0$ estimated by \citet{marconi04}.
The result of
this calculation is shown in Fig.\,\ref{fig1} for two different assumptions.
The dot-dashed curve shows the expected evolution if $\rho_{\rm BH}$ does not change
with redshift: the dashed curve shows the expected evolution if $\rho_{\rm BH}$
tracks the mass density in massive halos, $M_{\rm H} > 10^{11.5}$\,M$_{\odot}$,
calculated from the EPS mass function.  We see remarkably good agreement with
the evolution in bolometric luminosity density, derived from hard X-ray surveys
(solid curve and points with error bars),
if the mean radiative efficiency has a constant value 
$\left\langle\epsilon\right\rangle \simeq 0.04$.
This is a promising result:  there is no other free normalisation of the predicted
luminosity density, and the value of 
$\left\langle\epsilon\right\rangle$ required matches very well
both theoretical expectation ($\epsilon \la 0.06$ for accretion onto a Schwarzschild
black hole) and determinations based on a comparison of the X-ray background
and the local relic black hole mass density \citep[e.g.][]{marconi04}.

We can however also see that the prediction is too high at $z<0.5$, overproducing
the bolometric luminosity density by a factor 2 at $z=0$.  This implies that
black hole growth and halo/galaxy growth have decoupled by the present day.
This result may well be related to the phenomenon of downsizing noted
at $z=0$ by \citet{heckman04}.  In that work it appears that the mean Eddington
ratio becomes a function of mass, a result that also appears to be confirmed
by \citet{netzer06}.  Possible physical causes of decoupling could
be that either it is an apparent effect caused by a decrease in radiative
efficiency at low Eddington ratios \citep[e.g.][]{narayan95,beckert02}, or that 
black hole growth really does slow down as a result of the changing environments
of galaxies towards $z=0$, with an increasing prevalence of galaxies forming
into groups and clusters.

\section{Conclusions and further thoughts}
It seems that the hypothesis of Pure Coeval Evolution of black holes
and dark halos reproduces rather well the observed bolometric luminosity
density produced by AGN at $z<3$, implying that, broadly-speaking, black
holes and their hosts grow together.  The dramatic decrease in AGN activity
towards $z=0$ is thus seen to be a result of the slow-down in growth of galaxies
themselves.  This parallel evolution does not necessarily imply a direct causal
link between them, but it is attractive to invoke such a causality, especially
as this would also help explain the very tight $M-\sigma$ relation between black
holes and their hosts in the local universe \citep{tremaine02}.

One of the main successes of this hypothesis is that it reproduces the observed
bolometric luminosity density assuming a very reasonable value for the mean radiative
efficiency, $\langle\epsilon\rangle \simeq 0.04$ - no previous model of AGN
evolution has been able to predict the absolute value of the luminosity density.
It is worth noting that Pure Coeval Evolution produces a better match to the
luminosity density evolution than current generations of semi-analytic models
\citep[e.g.][]{croton06}.

A factor-two decoupling between black holes and galaxy/halo growth does seem to occur
at low redshift, more measurements of Eddington ratio as a function of black hole
mass, environment and redshift should enable us to distinguish alternative explanations
for this effect.

More observational and theoretical work is needed to further test the scenario
and to understand the physics of the coevolution process - it is not obvious
why the growth of dark-matter-dominated halos with total mass $\sim 10^{12}$\,M$_{\odot}$ should have
such a direct influence on the growth of central black holes with mass
$10^{6-8}$\,M$_{\odot}$.  
Meanwhile, other work in progress (Babi\'{c} et al., these proceedings
and in preparation) shows how both the AGN luminosity function and the X-ray
background can be successfully reproduced within this framework, and it would be very interesting
to extend this to higher redshifts.

A consequence of the predicted high Eddington ratios at high redshifts is that we expect
a large fraction of galaxies to have nuclear outflows from their growing black holes
\citep[e.g.][]{king03,kingpounds03} - searching for evidence of this would be important
both for testing the hypothesis and for helping to understand the importance of feedback 
when galaxies formed.

\acknowledgements %%% Text of acknowledgements runs on after this command.
I thank Y.\,Ueda for providing the data points used for the calculation shown
in Fig.\,\ref{fig1}.

%%% THE BIBLIOGRAPHY
%%%
%%% CONSULT SECTION 3 OF "INSTRUCTIONS FOR AUTHORS" FOR HOW TO USE NATBIB.
%%% AUTHORS ARE ENCOURAGED TO USE EITHER THE "THEBIBLIOGRAPY" ENVIRONMENT
%%% BY UNCOMMENTING (DELETING THE "%" SYMBOL) THE COMMANDS BELOW, OR BY
%%% USING THE BIBTEX ENVIRONMENT. TO FIND OUT WHICH IS APPLICABLE TO YOUR
%%% CONTRIBUTION, CONSULT THE VOLUME EDITORS FOR YOUR PROCEEDINGS.
%%%

\end{document}